\RequirePackage{lineno}

\pdfoutput=1
\documentclass[twocolumn,superscriptaddress,floatfix,preprintnumbers,amssymb,amsmath]{revtex4}

\ifx\pdfoutput\undefined
  \usepackage[dvips]{graphicx}
  \usepackage[dvips]{color}
\else
  \usepackage[pdftex]{graphicx}
  \usepackage[pdftex]{color}
\fi

\usepackage{dcolumn}
\usepackage{amsmath}
\usepackage{latexsym}

\usepackage{units}
\usepackage{ulem}

\begin{document}

\title{Photon-activated electron hopping in a single-electron trap enhanced by Josephson radiation}

\author{S.~V.~Lotkhov}
\affiliation{Physikalisch-Technische Bundesanstalt, Bundesallee 100, 38116, Braunschweig, Germany}
\author{B.~Jalali-Jafari}
\affiliation{Physikalisch-Technische Bundesanstalt, Bundesallee 100, 38116, Braunschweig, Germany}
\affiliation{Department of Microtechnology and Nanoscience (MC2), Chalmers University of Technology, SE-412 96 G\"oteborg, Sweden}
\author{A.~B.~Zorin}
\affiliation{Physikalisch-Technische Bundesanstalt, Bundesallee 100, 38116, Braunschweig, Germany}

\begin{abstract}

Using a Josephson junction interferometer (DC SQUID) as a microwave source for irradiating a single-electron trap, both devices fabricated on the same chip, we study the process of photon-assisted tunneling as an effective mechanism of single photon detection. High sensitivity down to a very small oscillation amplitude $v_{\rm J} \sim$~10~nV$\ll E_{\rm act} \lesssim hf_{\rm J}$ and down to low photon absorption rates $\Gamma_{\rm ph} \sim$~(1--50)Hz, as well as a clear threshold type of operation with an activation energy $E_{\rm act} \sim$~400~$\mu$eV, are demonstrated for the trap with respect to the microwave photons of frequency $f_{\rm J} \sim$~(100--200)GHz. Tunable generation is demonstrated with respect to the power and frequency of the microwave signal produced by the SQUID source biased within the subgap voltage range. A much weaker effect is observed at the higher junction voltages along the quasiparticle branch of the $I-V$ curve; this response mostly appears due to the recombination phonons.

\end{abstract}


\maketitle

Operation of single-electron tunneling (SET) circuits is known to be significantly influenced by microwave radiation coupled to tunnel junctions \cite{IngoldNazarov}. The related phenomena are regularly observed in experiments in the form of noise-induced charge tunneling, and so-called photon-assisted tunneling (PAT) \cite{MartinisNahum} mechanisms have been extensively studied over the last two decades, both in the normal conducting \cite{Keller1998,Covington2000} and superconducting \cite{Hergenrother1995,Pekola2010,Saira2010,Barends2011,Lotkhov2011,Lotkhov2012} systems with tunnel junctions. A straightforward demonstration of PAT in an electron pump has been provided in Ref.~\cite{Covington2000}, where an external source of the high-frequency signal was used. Recently, several studies have been performed which involve compact experimental arrangements combining on the same chip a tunnel-junction-based source of the microwaves with a strongly coupled (also tunnel-junction-based) photon detector \cite{Billangeon,Lotkhov2012}. On the other hand, a dramatic reduction of the background PAT rates has been achieved with the help of a double-shielded cryogenic sample holder \cite{Kemppinen2011} or  on-chip line filtering \cite{Lotkhov-LT26} employed for the tunneling experiments.

\begin{figure}[t]
\centering%
\includegraphics[width=1.0\columnwidth]{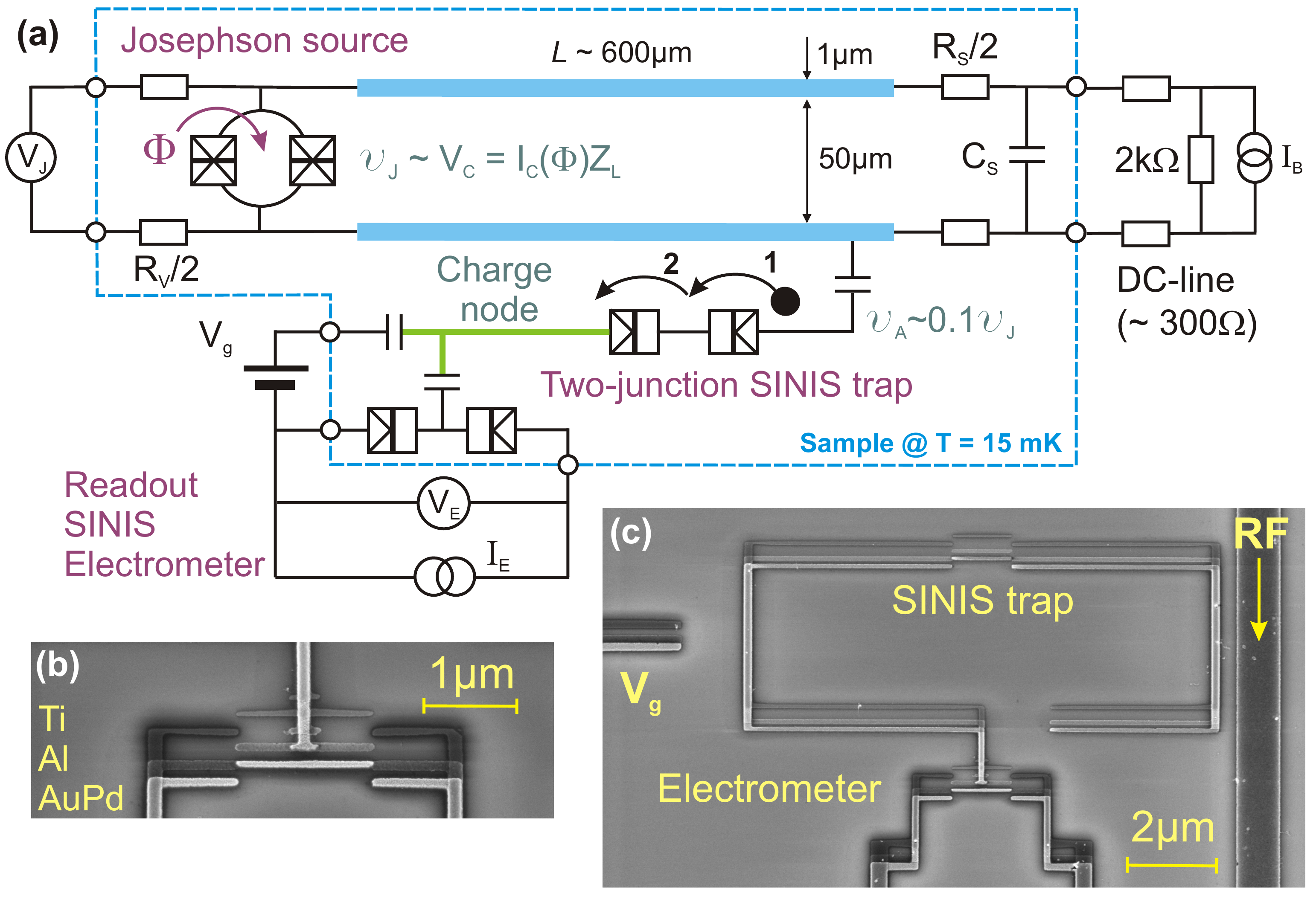}
\caption{
(Color online) (a) Electrical circuit diagram of the experimental device. In the Josephson part of the circuit, the normal resistance of the SQUID is $R_{\rm J} \approx$~2.3~k$\Omega$ and the maximal critical current is $I^{\rm max}_{\rm C} \equiv I_{\rm C} (\Phi = 0) \approx$~65~nA. The resistances $R_{\rm V} \approx$~1~k$\Omega$ and $R_{\rm S} \approx$~240~$\Omega$ are realized as bilayer microstrips of AuPd/Ti \cite{JalaliJafari2014} with the square sheet resistance $\rho \approx$~80~$\Omega$. The line termination includes an interdigital capacitor, $C_{\rm S} \sim$~0.1~pF, with a low impedance at high frequencies, $Z_{\rm C}(f\sim$100~GHz) $\sim$~10~$\Omega$. (b) Blow-up image of the SINIS electrometer. The materials of the three replicas seen in the graph are Ti (30~nm, used for resistive low-pass filtering inserts into the DC leads, about 20~$\mu$m apart from the junctions, not shown), Al (20~nm, oxidized $in~situ$ to make a tunnel barrier) and AuPd (30~nm) for the SIN junctions. (c) SEM image of the SINIS detector (a double-junction SINIS trap coupled to a SINIS electrometer) coupled capacitively to the waveguide. The same nominal area 20$\times$20~nm$^2$ is e-beam-written for all junctions (see Ref.~\cite{Lotkhov2011} for the details of the fabrication). The test resistance of the electrometer is $R_{\rm T} \approx$~1.2~M$\Omega$, the charging energy $E_{\rm C} \equiv e^2/2C_\Sigma \approx$~180~$\mu$eV, where $C_\Sigma$ is the total capacitance of the small  intermediate island.
}
\label{Fig1}
\end{figure}

In this Letter, we report on the direct observation of the photon activated tunneling of charge in a hybrid Josephson-SET circuit, integrating on the same chip a tunable microwave source and an SET-based single-photon detector coupled to this source via a superconducting coplanar waveguide. The experimental layout is shown in Fig.~\ref{Fig1} and includes a DC-SQUID-based Josephson oscillator, a two-wire transmission line and a two-junction charge trap read out by an SET electrometer \cite{Lotkhov2011}. We demonstrate a strong effect of controllable microwave irradiation on the switching statistics in the trap. A comparison of the experimental data with the PAT theory \cite{IngoldNazarov,MartinisNahum} clearly demonstrates a quantum nature of interaction between the weak electromagnetic wave and the electron tunneling system. In contrast to our previous experiment reported in Ref.~\cite{Lotkhov2012}, the detector is placed clearly apart from the source, so that the possible acoustic (phonon-mediated) component of the signal (see e.g. Ref.~\cite{Schinner2009Gasser2010}) reaching our detector is deliberately reduced.

The oscillations amplitude and thus the microwave power generated by the SQUID source is designed in such a way that it can be varied by applying a magnetic field $B \sim~ \Phi_0/A \approx$~5~mT, where $\Phi_0 \approx $~2.07$\times 10^{-15}$~Wb is a flux quantum and $A \approx$~0.4~$\mu$m$^2$ is a loop area of our device. The junctions, 0.2$\times$0.24~$\mu$m$^2$ in size, are made of aluminum (see Ref.~\cite{JalaliJafari2014} for the fabrication details) and biased via the miniature resistors as shown in Fig.~\ref{Fig1}. DC shunting was avoided in order to reduce on-chip generation of heat and thus to keep the thermal widening of the generation linewidth $\delta\Gamma \sim$~1~GHz \cite{JalaliJafari2014} small. The impedance of biasing circuitry seen by the SQUID at Josephson frequency, $R \approx \left(R_{\rm V}^{-1}+Z_{\rm L}^{-1} \right)^{-1}$, is that of the parallel connection of the isolating resistors $R_{\rm V}\approx$~1~k$\Omega$ and the specific impedance of the transmission line $Z_{\rm L} \approx$~260~$\Omega$ terminated by a matched load, as discussed below.

As shown in our previous work \cite{JalaliJafari2014}, one can produce voltage oscillations of the fundamental Josephson frequency $hf_{\rm J} = 2eV_{\rm J}$ by biasing the junctions to sufficiently large Josephson voltages within the subgap range, $V_{\rm J} < 2\Delta/e \approx$~400~$\mu$V, where $\Delta$ is the superconducting gap of Al. For the parameters of our present device (see caption to Fig.~\ref{Fig1}) and at zero magnetic flux, $\Phi = 0$, this frequency range extends to 50~GHz~$<f_{\rm J}<$~200~GHz and the amplitude of the lead harmonics is expected to vary correspondingly in the range 13~$\mu$V$>v_{\rm J}>$8.8~$\mu$V, being on the same scale as the critical voltage $V_{\rm C} \equiv I_{\rm C}R \approx $13.7~$\mu$V. A much lower amplitude is estimated for the higher harmonics (see Ref.~\cite{JalaliJafari2014} for more details), thus making the signal quasi-monochromatic. For non-zero flux, the SQUID modulation of the critical current $I_{\rm C}$ results in the proportional variation of the amplitude, $v_{\rm J} \propto I_{\rm C}(\Phi)$ \cite{LikharevRUS}.

The microwave signal produced by the Josephson source is delivered to the detector via a coplanar waveguide fabricated from Al in the same layer as the SQUID. In order to achieve a high specific impedance $Z_{\rm L}$, and thus to increase the oscillation amplitude $v_{\rm J} \sim I_{\rm C}Z_{\rm L}$, this coupling element was implemented as a two-wire transmission line.  As the length of the line is on the same scale as the radiation wavelength, $\lambda \sim$~1.2~mm at $f\sim$~100~GHz, a matched $R_{\rm S}-C_{\rm S}$ termination was included in the layout, as shown in  Fig.~\ref{Fig1}, in order to prevent resonances and provide a flat frequency dependence for the generated and transmitted microwave power.

\begin{figure}[t]
\centering%
\includegraphics[width=0.9\columnwidth]{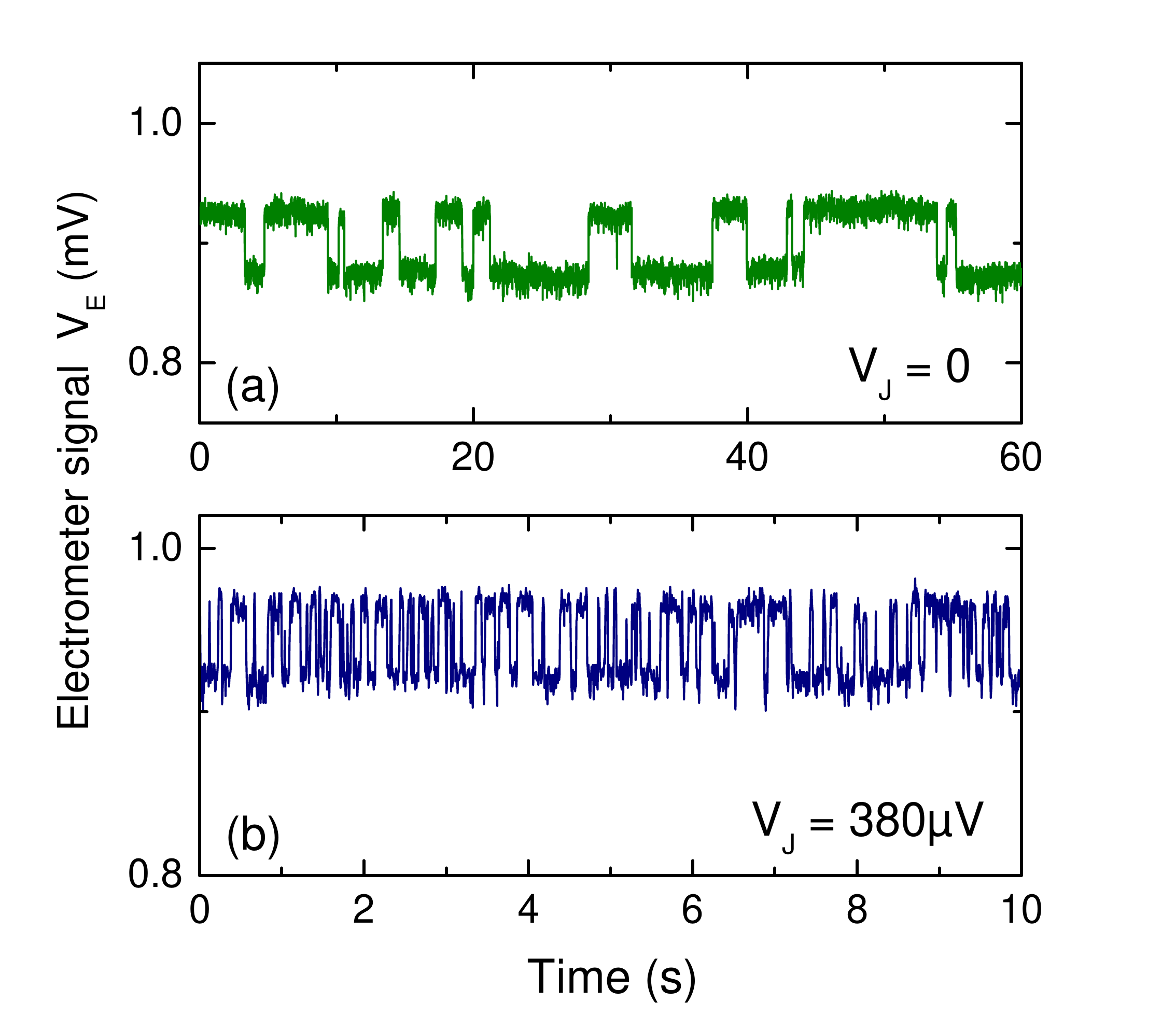}
\caption{
(Color online) Switching traces of the SINIS trap recorded (a) without irradiation and (b) under irradiation from the Josephson source tuned to a reduced value of $I_{\rm C} \approx$~15~nA. To minimize a possible backaction of the SINIS electrometer to the trap, its bias current is set to a low value, $I_{\rm E} \approx$~40~pA. Note the different scales of the time axis in the upper and bottom panels! The rate $\Gamma_{\rm ph}$ can be derived from this random switching trace as an average number of switchings per unit time.
}
\label{Fig2}
\end{figure}

For microwave detection, we used a double-junction SINIS-type SET trap with the activation barrier $E_{\rm act}$, enhanced by the superconductivity ("S") of the (floating-potential) side terminals connected to the small normal conducting  ("N") island via oxidation barriers ("I") \cite{Lotkhov2011,Lotkhov2012}. One of the terminals, the "charge node" depicted in Fig.~\ref{Fig1}(a), is capacitively coupled to an SET electrometer (also of SINIS type, as it is made in the same layer as the trap), to monitor the lifetime statistics of the trapped charge states. Charge transfer across the double junction of the trap involves two consecutive tunneling steps numbered in Fig.~\ref{Fig1}. The rate of the first tunneling step $\Gamma_1$ is low due to the negative energy gain $E_1 = -E_{\rm act} = -\Delta -  E_{\rm Q1}$, where $E_{\rm Q1}$ is a Coulomb energy required for charging up a small N-island by one electron in the step 1. The rate of the second tunneling step $\Gamma_2$ is related to a more favorable energy gain $E_2 = -\Delta +  E_{\rm Q2}$, where $E_{\rm Q2}$ is a Coulomb energy released by discharging the N-island in the step 2. For simplicity, we tune the gate voltage $V_{\rm g}$ to achieve the symmetry point $E_{\rm Q1} \approx E_{\rm Q2}$. Except for particular "degenerate" points where $E_{\rm Q1} \approx E_{\rm Q2} \approx 0$, the second rate is much higher than the first one, $\Gamma_2 \gg \Gamma_1$. In experiment, the second step is typically not time resolved by the DC electrometer used and the tunneling sequence appears as a single charge hopping event over the barrier $E_{\rm act}$ at a rate $\Gamma_{\rm ph} \approx \Gamma_1/2$. The factor $\times$1/2 appears beside $\Gamma_1$ due to 1/2 probability of the backward tunneling in the second step through the first junction without producing an electrometer signal.

At a low temperature, $k_{\rm B}T \ll E_{\rm act}$, the rate of the spontaneous hopping is low, typically $\Gamma_{\rm ph}^0 \sim$~(0.001--1)~Hz for $E_{\rm act}/h >$~100~GHz, and is mostly related to the absorption of background photons which appear due to microwave leaks into the sample cavity \cite{Lotkhov2011,Kemppinen2011}. Applying the microwave signal $hf_{\rm J} = 2eV_{\rm J} > E_{\rm act}$, we observe a dramatic increase of the switching intensity (up to several orders of magnitude), as shown in Fig.~\ref{Fig2} for the symmetry point featured by the equal lifetimes of the two alternating states of the charge node.

\begin{figure}[t]
\centering%
\includegraphics[width=0.95\columnwidth]{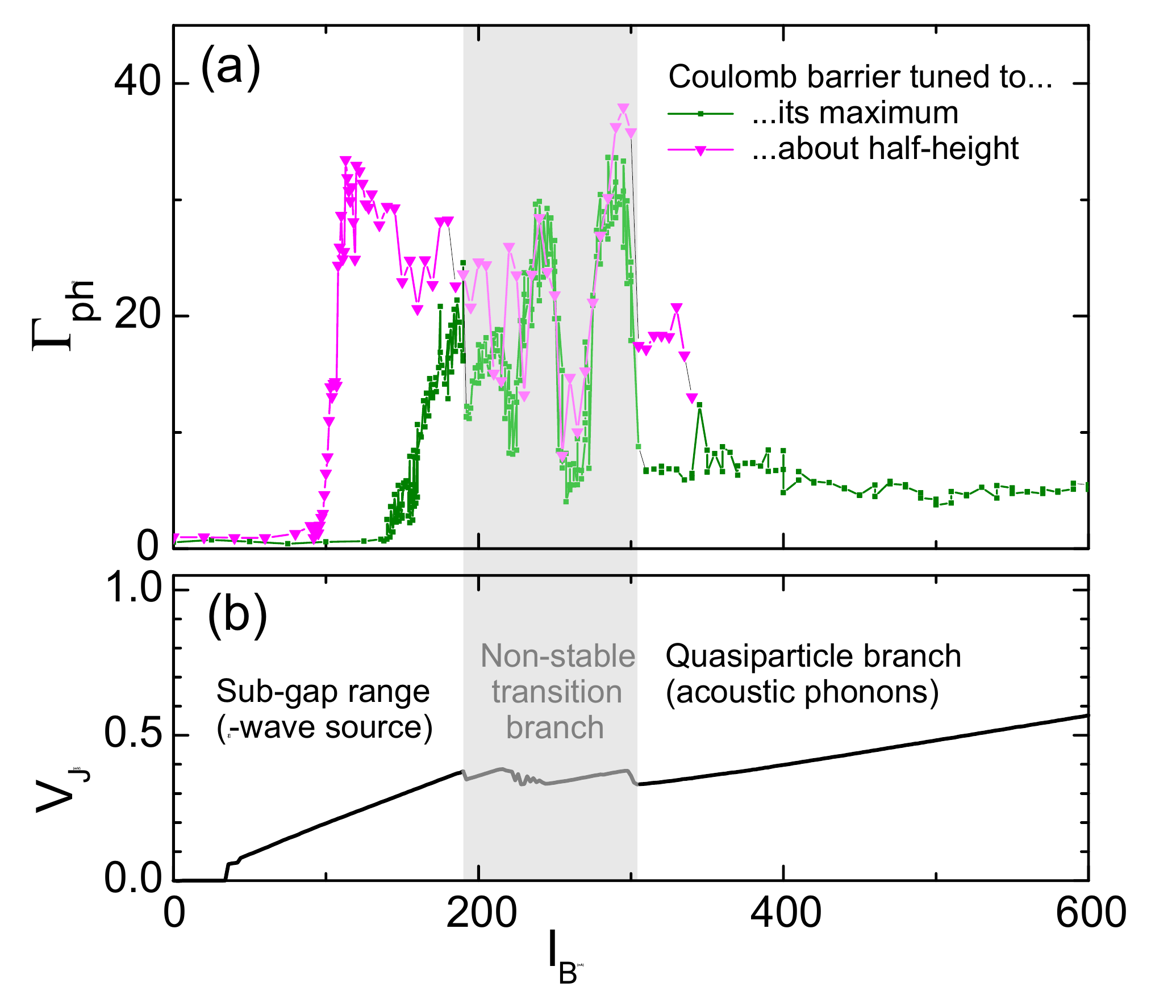}
\caption{
(Color online) (a) Photon-activated rate $\Gamma_{\rm ph}$ vs. total bias current $I_{\rm B}$ for two different heights of the trapping barrier \cite{Ec} and an intermediate value of $I_{\rm C} \approx$~24~nA. (b) $I-V$ curve of the source measured under the same conditions as those used to irradiate the detector in panel (a) (for the biasing circuit, see Fig.~\ref{Fig1}).
}
\label{Fig3}
\end{figure}

The statistics of state switchings is studied in more detail by biasing the source to the different operating regimes shown in Fig.~\ref{Fig3}(b). In the sub-gap range of the voltage measurement $V_{\rm J} <2\Delta/e$, no response is observed around the supercurrent branch of the Josephson device, followed by a sharp rise in the signal at the voltage threshold dependent on the height of the Coulomb barrier in the trap, see Fig.~\ref{Fig3}(a). In the transition section of the plot, marked as a grey area, both the $I-V$ curve and the detector's response exhibit strong irregularities which can be explained by overheating effects in the Josephson junctions. Here, considerable thermal suppression is expected for the critical current $I_{\rm C}$, leading to strong variations of the radiated microwave power. For above-gap voltage values, $V_{\rm J} \gtrsim 2\Delta$ (the high current values, $I_{\rm B} >$~300~nA), i.e., along the quasiparticle branch of the junction $I-V$ curve, only a weak response is registered, thus indicating that vanishing microwave power and only a weak phonon-mediated signal is provided to the detector in this regime (the latter is discussed in more detail below).

\begin{figure}[t]
\centering%
\includegraphics[width=1.0\columnwidth]{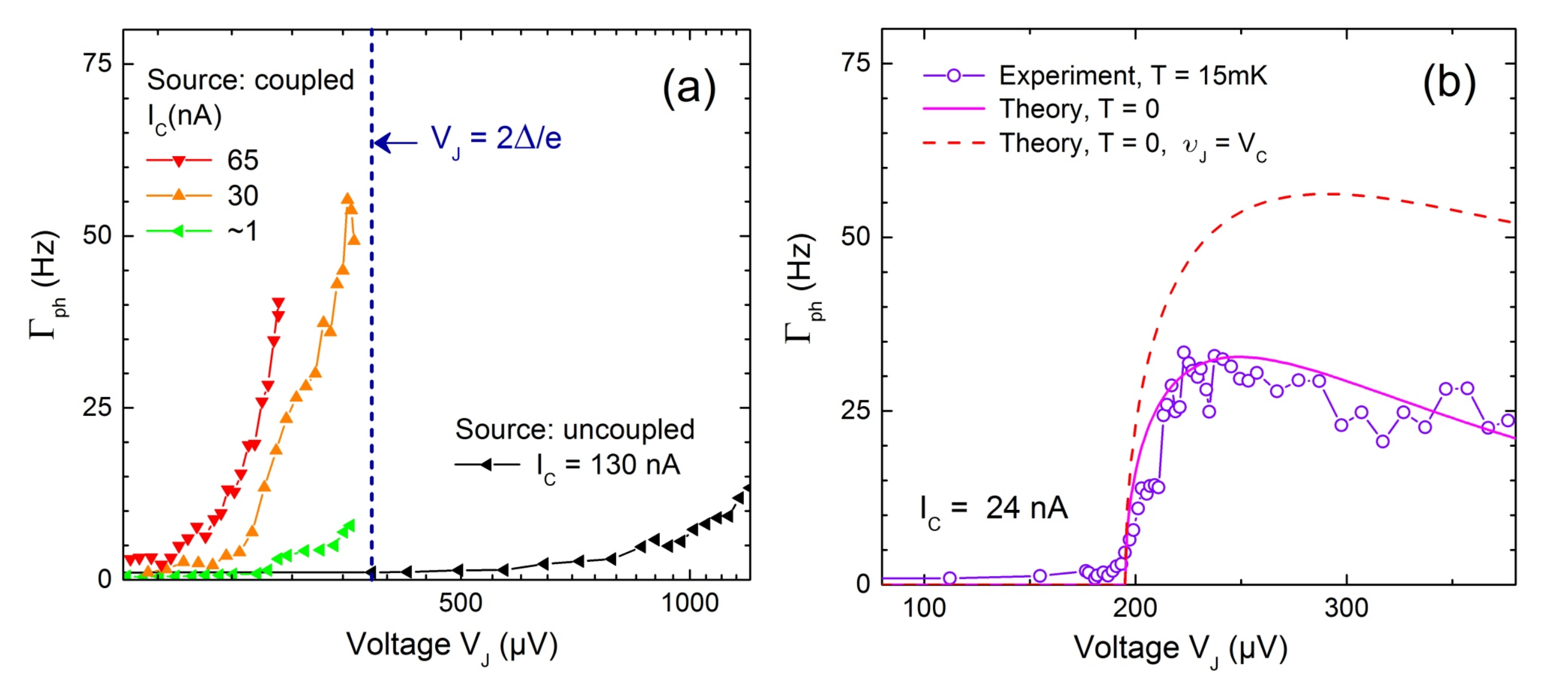}
\caption{
(Color online) The rate $\Gamma_{\rm ph}$ as a function of average voltage $V_{\rm J}$ plotted (a) for several values of the critical current $I_{\rm C}$ in the $coupled$ SQUID generator as well as for the maximum, twice as large value of $I'_{\rm C}$ in the $uncoupled$ source and (b) for the conditions of the experiment shown in Fig.~\ref{Fig3}, within the stable voltage range 80~$\mu$V~$\lesssim V_{\rm J}\lesssim$~380~$\mu$V. The solid line shows a theoretical fit using Eq.~\ref{analytic} and assuming $R_{\rm N} = 1/2R_{\rm T}$ as well as applying the activation energy $E_{\rm act} =$~390~$\mu$eV and the coupling factor $\alpha \equiv v_{\rm A}/v_{\rm J} =$~0.12 as fitting parameters. The dashed line shows the rate $\Gamma_{\rm ph}$ calculated for a fixed oscillation amplitude of the source, $V_{\rm C} \approx$~5~$\mu$V, thus illustrating the frequency response curve for the SINIS detector.
}
\label{Fig4}
\end{figure}

Figure~\ref{Fig4} shows the dependence of the switching rate $\Gamma_{\rm ph}$  on the average voltage across the junction $V_{\rm J}$ which is the measure of the oscillation frequency $f_{\rm J}$ and the photon energy $E_{\rm ph} = 2eV_{\rm J}$.
Panel (a) demonstrates the power tunability of the SQUID generator by suppressing the critical current from its maximum value $I_{\rm C} \approx$~65~nA --- for which the response curve $\Gamma_{\rm ph} - V_{\rm J}$ exhibits a steep rise up to the levels beyond the frequency domain of representative statistics, $\Gamma_{\rm ph} \lesssim$~50~Hz --- down to its tunable minimum of about 1~nA. The latter curve, i.e., that corresponding to $I_{\rm C} \sim$~1~nA, indicates, besides the substantial count reduction down to the "dark" level, $\Gamma_{\rm ph} \sim$~ 1~Hz, also a resolved response even to very weak microwave oscillations across the SINIS double junction with the estimated amplitude as small as $v_{\rm A} \sim 0.1v_{\rm J} \sim$~10~nV. This scale lies many orders below the activation threshold, $ev_{\rm A} \lll E_{\rm act} (\sim$~400~$\mu$eV)$\lesssim hf_{\rm J}$, which is a clear manifestation of the quantum nature of the microwave-to-charge interaction observed.

To evaluate the fraction of the response signal which appears due to the recombination phonons created by the quasiparticle component of the tunnel current, an uncoupled (i.e., without transmission line coupling) reference source of the same type was placed on the substrate at a similar distance from the detector, $L =$~600~$\mu$m. Despite a larger value of $I'_{\rm C} \approx$~130~nA (i.e., four times (!) higher irradiation power), no response was registered in the sub-gap voltage range of $V_{\rm J}$; this proves an efficient electromagnetic decoupling of the uncoupled source, on the one hand, and strong sub-gap suppression of the quasiparticle leak current, on the other hand. Along the quasiparticle branch of the $I-V$ curve, $V_{\rm J} > 2\Delta$, see Fig.~\ref{Fig3}(b), a weak parabolically-shaped increase of $\Gamma_{\rm ph}$ is observed, see Fig.~\ref{Fig4}(a), which corresponds to an approximately linear power dependence of the signal and to an estimated efficiency of one detected phonon out of $\sim 10^{12}$ which are generated in the source. To compare: the microwave detection efficiency of the coupling layout is 1 photon out of $\sim 10^8$ created by the source, which, however, reflects the fact that the generated gross microwave power is almost completely dissipated in the biasing circuitry with the resistors $R_{\rm S}$ and $R_{\rm V}$.

We find it useful to compare the experimental data with a theory and, in particular, to obtain in this way the value of the activation energy $E_{\rm act}$ and the frequency response curvature; both of these properties are important characteristics of the SINIS detector. Figure~\ref{Fig4}(b) shows the comparison of the lower-barrier data from Fig.~\ref{Fig3}(a), now plotted as a function of voltage $V_{\rm J}$, with the calculation results based on the PAT model described in Ref.~\cite{MartinisNahum} and developed on the basis of the standard $P$-theory of electron tunneling in the presence of an electromagnetic environment \cite{IngoldNazarov}. The details of the calculation are presented in Ref.~\cite{JalaliJafari2015}. In the zero-temperature approximation, assuming a narrow linewidth of the Josephson oscillations, $\delta\Gamma \to 0$, and a non-dissipative environment of the SINIS trap, we obtain the following analytical expression for the PAT rate $\Gamma_{\rm ph}$ in the symmetry point:

\begin{equation}
\Gamma_{\rm ph} \approx \frac{\Gamma_1}{2} \approx \frac{\Delta}{64R_{\rm N}} \left[\frac{v_{\rm A}}{2eV_{\rm J}}\right]^2\sqrt[]{\left(\frac{2eV_{\rm J}-E_{\rm act}}{\Delta}+1\right)^2-1},
\label{analytic}
\end{equation}
where $R_{\rm N}$ is the tunnel resistance of one junction in the (uniform) SINIS double junction. The result of the calculation is shown in Fig.~\ref{Fig4}(b) and provides the expectable fitting values for the threshold energy $E_{\rm act}$  \cite{Ec} and the coupling factor $\alpha$ as judged based on the details of the layout. The dashed line in Fig.~\ref{Fig4}(b) represents the frequency dependence of the detector's response: the rate $\Gamma_{\rm ph}$, calculated as a function of voltage $V_{\rm J}$ for a fixed value of the microwave amplitude $v_{\rm J} = V_{\rm C} \approx$~5~$\mu$V (which is the maximum for the present, reduced value of $I_{\rm C} =$~24~nA). Here, we assume a non-lossy behavior of the transmission line at a low magnetic field, $B \approx$~1.8~mT,  applied to the SQUID. The shape of the curve is typical for the threshold type of detection, and also demonstrates an almost-flat response characteristics above the threshold, 120~GHz~$<f_{\rm J}<$~190~GHz.

Finally, we note that, by varying the irradiation power over a wide accessible range, it is possible to estimate the dynamic range of the SINIS detector with respect to the measurable counting rates. The present, relatively narrow range, $\Gamma_{\rm ph}^0(\lesssim$1~Hz~)$\lesssim \Gamma_{\rm ph} \lesssim$~50~Hz, is limited by the non-deterioration condition for the counting statistics. In particular, we try to avoid the contribution of the microwave background at low frequencies and the statistical drop-out of the shorter lifetimes at the upper cutoff frequency. The latter is set by the time resolution of the DC electrometer biased by a low, pA-range current. However, considerable improvements should be expected by implementing a higher degree of microwave shielding of the sample cavity \cite{Kemppinen2011}, for example, or by using a faster electrometry (see e.g. Ref.~\cite{Ferguson2006} and the citations therein).

To conclude, using a hybrid Josephson-SET on-chip circuit, we investigated the photon-assisted single-electron tunneling rates enhanced by Josephson microwave radiation. The counting statistics was demonstrated to be governed by the quantum interaction of the microwaves in the frequency range 100~GHz~$\lesssim f_{\rm J} \lesssim$~200~GHz with an SET trap as a discrete tunneling system. Tunable operation of the Josephson source was demonstrated with respect to the microwave power and frequency. Reasonable agreement with theory was observed.

We acknowledge experimental support from T. Weimann and V. Rogalya. This work was funded by the Joint Research Project MICROPHOTON. Joint Research Project MICROPHOTON belongs to the European Metrology Research Programme (EMRP). The EMRP is jointly funded by the EMRP participating countries within EURAMET and the European Union.

\end{document}